\documentclass[sigconf]{acmart}

\copyrightyear{2023}
\acmYear{2023}
\setcopyright{acmlicensed}\acmConference[WSDM '23]{Proceedings of the Sixteenth ACM International Conference on Web Search and Data Mining}{February 27-March 3, 2023}{Singapore, Singapore}
\acmBooktitle{Proceedings of the Sixteenth ACM International Conference on Web Search and Data Mining (WSDM '23), February 27-March 3, 2023, Singapore, Singapore}
\acmPrice{15.00}
\acmDOI{10.1145/3539597.3570466}
\acmISBN{978-1-4503-9407-9/23/02}

\graphicspath{{figs/}}

\usepackage{xcolor}
\usepackage{xspace}
\usepackage{enumitem}
\newtheorem{definition}{Definition}

\newcommand{\app}{WeChat\xspace}
\newcommand{\ours}{SGRL\xspace}

\begin{document}

\title[SGRL for Black Market Account Detection]{Self-supervised Graph Representation Learning for\\Black Market Account Detection}

\author{Zequan Xu}
\authornote{The first two authors contributed equally. Work done when they were interns at Tencent Inc.}
\affiliation{%
 \institution{School of Informatics\\Xiamen University}
 \city{Xiamen}
 \state{Fujian}
 \country{China}
}
\email{xuzequan@stu.xmu.edu.cn}

\author{Lianyun Li}
\authornotemark[1]
\affiliation{%
 \institution{School of Informatics\\Xiamen University}
 \city{Xiamen}
 \state{Fujian}
 \country{China}
}
\email{lilianyun@stu.xmu.edu.cn}

\author{Hui Li}
\authornote{Corresponding author.}
\affiliation{%
  \institution{School of Informatics\\Xiamen University}
  \city{Xiamen}
  \state{Fujian}
  \country{China}
}
\email{hui@xmu.edu.cn}

\author{Qihang Sun}
\affiliation{%
  \institution{Tencent Inc.}
  \city{Guangzhou}
  \state{Guangdong}
  \country{China}
}
\email{aaronqhsun@tencent.com}

\author{Shaofeng Hu}
\affiliation{%
  \institution{Tencent Inc.}
  \city{Guangzhou}
  \state{Guangdong}
  \country{China}
}
\email{hugohu@tencent.com}

\author{Rongrong Ji}
\affiliation{%
  \institution{School of Informatics\\Xiamen University}
  \city{Xiamen}
  \state{Fujian}
  \country{China}
}
\email{rrj@xmu.edu.cn}

\renewcommand{\shortauthors}{Xu et al.}

\begin{abstract}
  Nowadays, Multi-purpose Messaging Mobile App (MMMA) has become increasingly prevalent. MMMAs attract fraudsters and some cybercriminals provide support for frauds via black market accounts (BMAs). Compared to fraudsters, BMAs are not directly involved in frauds and are more difficult to detect. This paper illustrates our BMA detection system SGRL (Self-supervised Graph Representation Learning) used in WeChat, a representative MMMA with over a billion users. We tailor Graph Neural Network and Graph Self-supervised Learning in SGRL for BMA detection. The workflow of SGRL contains a pretraining phase that utilizes structural information, node attribute information and available human knowledge, and a lightweight detection phase. In offline experiments, SGRL outperforms state-of-the-art methods by 16.06\%-58.17\% on offline evaluation measures. We deploy SGRL in the online environment to detect BMAs on the billion-scale WeChat graph, and it exceeds the alternative by 7.27\% on the online evaluation measure. In conclusion, SGRL can alleviate label reliance, generalize well to unseen data, and effectively detect BMAs in WeChat.
\end{abstract}



\maketitle


\section{Introduction}
\label{sec:intro}

Nowadays, one single mobile app tends to integrate multiple functionalities. 
Such \underline{M}ulti-purpose \underline{M}essaging \underline{M}obile \underline{A}pps (MMMAs) bring convenience to mobile users. \app\footnote{\url{https://www.wechat.com/en}} is a representative MMMA with over
a billion users. One can use \app for text/voice
messaging and voice/video calls, login to mobile games, or use the digital payment service to directly transfer money to other users.

\begin{figure}[t]
\centering
\includegraphics[width=0.99\columnwidth]{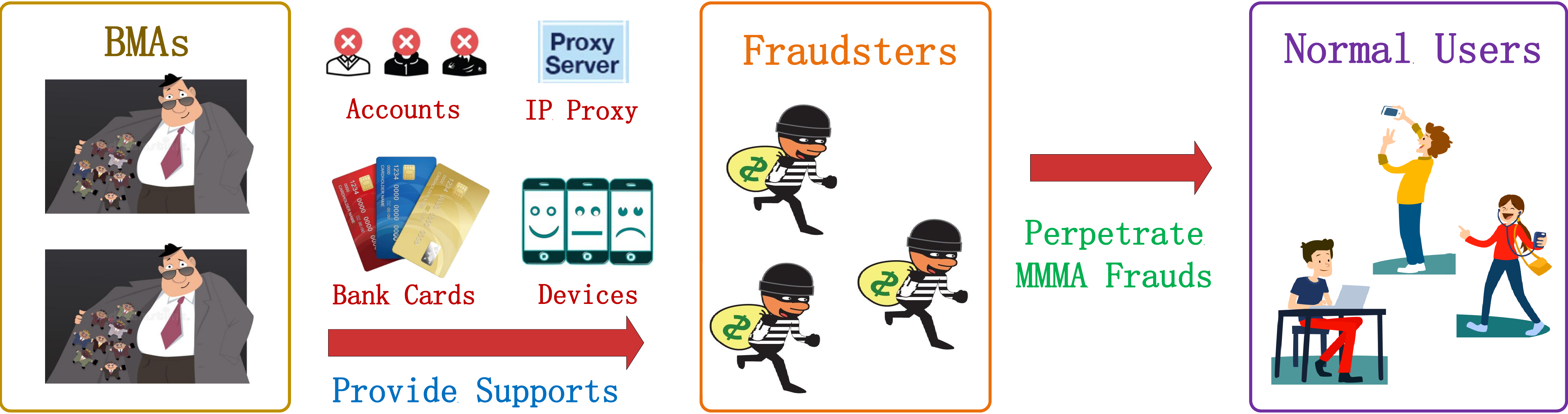}
\vspace{-10pt}
\caption{The role of black market accounts in \app.}
\label{fig:blackmarket}
\vspace{-10pt}
\end{figure}

The great convenience brought by \app also attracts cybercriminals~\cite{XuLSLLSH22}. They
deceive normal users through socializing over \app. 
After gaining trust, cybercriminals recommend stocks or
illegal gambling so that normal users transfer money to cybercriminals' \app accounts or reveal their passwords to cybercriminals. 
Since malicious
socialization and payment take place in one single app, the time left for normal
users to realize the fraud is limited. 
The massive profit brought by MMMA frauds expedites the development of the
\emph{black market} that supports frauds. As depicted in
Fig.~\ref{fig:blackmarket}, certain cybercriminals create \app accounts to sell
bulk quantities of \app accounts, bank accounts, phone numbers,
national identification numbers, devices, or IP proxies that can be
used to perpetrate frauds. 
In this paper, \emph{black market accounts} (BMAs) refer to such \app accounts that are not directly involved in frauds but provide support services for frauds.  
BMAs are the basis of MMMA frauds. Detecting and banning BMAs can significantly
increase the difficulty of perpetrating MMMA frauds.

\begin{figure*}[!t]
\centering
\includegraphics[width=0.99\textwidth]{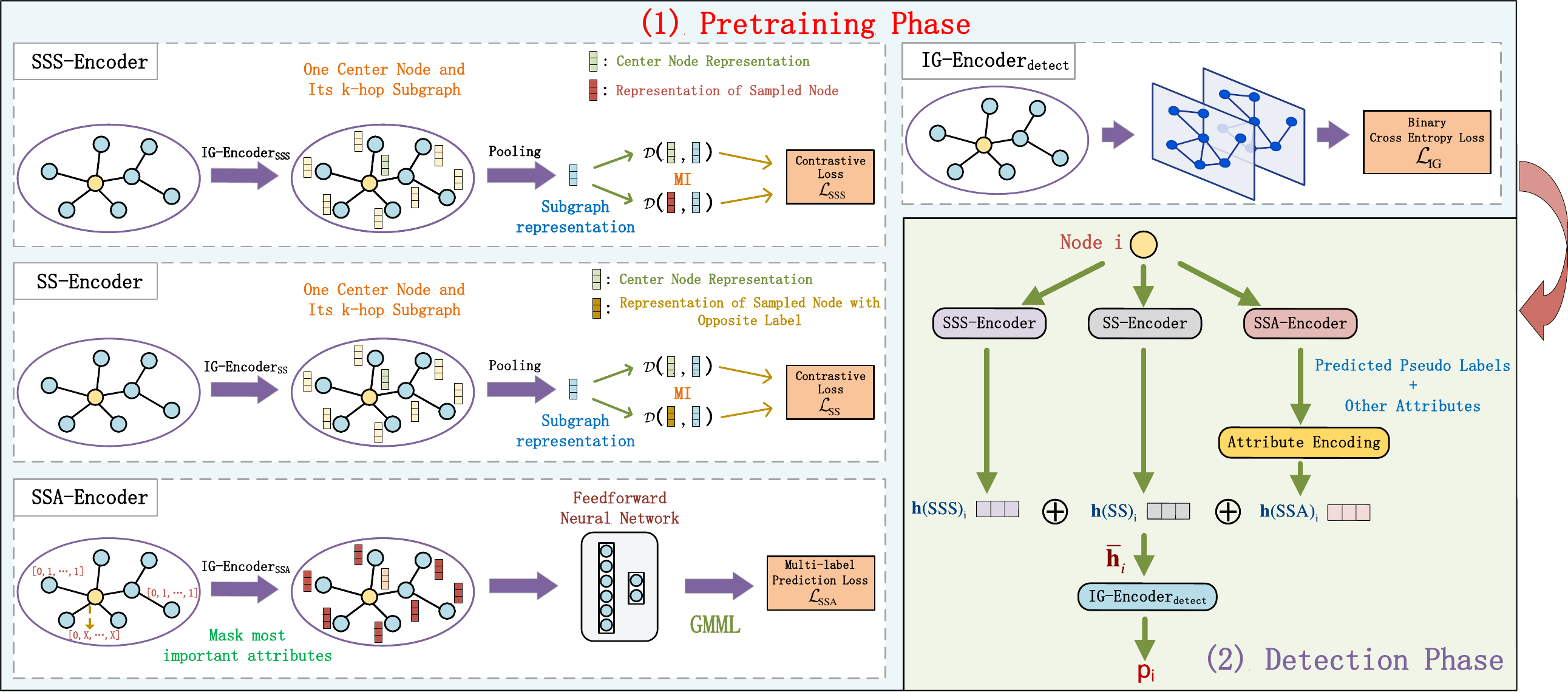}
\vspace{-10pt}
\caption{Overview of \ours. It consists of a pretraining phase and a detection phase.}
\label{fig:overview}
\end{figure*}

This paper illustrates our designed system for the 
\underline{B}M\underline{A} \underline{D}etection task
(the \emph{BAD} task) in \app. 
Due to the social nature of
\app, we construct a user-user interaction graph and conduct the BAD task. 
This way, the BAD task is closely related to
 Graph-based Anomaly Detection
(GBAD). 
However, \emph{the BAD task in \app has unique properties and it is more challenging than other GBAD tasks:} 

\vspace{5pt}
\noindent\textbf{P1: Huge and diverse data.} Massive users with diverse behaviors are using \app. 
It is a time-consuming and human-intensive process to manually identify BMA patterns. 
Even if a few patterns are found, using them in rule-based detection may not provide promising results: the diversity of \app data (e.g., interactions with some patterns only occur between certain users) makes it unlikely that these patterns are ubiquitous in the large graph.

\vspace{5pt}
\noindent\textbf{P2: Demand for generalization.} New users join \app every day. Hence, the detection system should adapt properly to unseen accounts. On the other hand, the large \app data requires a strong generalization ability: training should be offline on \emph{small} data and later the system can be used online for detecting BMAs over the \emph{large} \app graph to reduce the cost at detection time.

\vspace{5pt}
\noindent\textbf{P3: Lack of labels and great difficulty of manual labeling due to users' privacy.} Unlike financial fraud/risk detection that uses financial records after the amortization period as labels~\cite{XuSSACC21} or MMMA fraudster detection that uses victim reports as labels, labels for the BAD task are mostly \emph{unavailable}. 
Moreover, since BMAs do not directly get involved in frauds, manual labeling is \emph{difficult}, let alone its high cost: (1) 
The behavior types (e.g., chat and transactions) and behavior frequency of some BMAs, are not much different compared to other accounts. 
(2) Private information like chat content may help increase the accuracy of BMA labeling, but it is unaccessible in order to protect user privacy.

\vspace{5pt}
To combat with BMAs in \app, we develop SGRL, a \underline{S}elf-supervised \underline{G}raph \underline{R}epresentation \underline{L}earning based BMA detection system. 
We summarize our contributions as follows:

\vspace{5pt}
\noindent\textbf{C1: An Inductive Graph Neural Network (GNN) encoder for BMA representation learning.} To cope with P1 and P2, we adopt an Inductive GNN encoder (IG-Encoder) as the backbone of \ours to automatically capture rich information in \app graph and reduce the burden of BMA pattern investigation.

\vspace{5pt}
\noindent\textbf{C2: Novel structural/attribute encoders based on Graph Self-supervised Learning (GSSL).} To overcome P3, 
we explore GSSL~\cite{abs-2103-00111} to alleviate the reliance on BMA labels and design two types of encoders for modeling structures and node attributes, two essential parts in graph data, respectively:
\begin{itemize}[leftmargin=10pt,topsep=1pt,itemsep=0.3pt]
\item\textbf{C2.1: Contrastive structural encoders.} We observe that the surrounding structure (i.e., subgraph) of nodes is important in the BAD task. Thus, we design two contrastive structural encoder SSS-Encoder and SS-Encoder to encode the subgraph as a context into node representation. 

\item\textbf{C2.2: A self-supervised node attribute encoder.} Some node attributes are highly correlated with node labels. Hence, we design a node attribute encoder (SSA-Encoder) and use attributes as supervision. It captures attribute correlations through our designed Graph-based Multi-label prediction task with Missing Labels (GMML), which benefits the downstream BAD task.

\end{itemize}

\vspace{5pt}
GSSL also helps \ours handle P2: representation learning and detection are decoupled in \ours, and offline trained \ours on small data can be used for the online BAD task on the large graph.

\vspace{5pt}
\noindent\textbf{C3: An industrial BMA detection system deployed in \app.} We deploy \ours in \app with over a billion users. Offline and online experiments based on the large-scale MMMA data of \app show that \ours has better detection performance than existing methods and it can block the supports from BMAs for frauds.


\section{Our System \ours}
\label{sec:method}

\subsection{Overview}
\label{sec:overview}

We first construct a user-user interaction graph and extract node features from \app (Sec.~\ref{sec:prepare}). The graph is used to detect BMAs. 
As depicted in Fig.~\ref{fig:overview}, the workflow of \ours consists of two phases: pretraining and detection. The core part of \ours contains four encoders: SSS-Encoder, SS-Encoder, SSA-Encoder and IG-Encoder$_{\text{detect}}$.
They are constructed based on an inductive GNN encoder IG-Encoder (Sec.~\ref{sec:gnn}). 
SSS-Encoder and SS-Encoder are contrastive structural encoders (Sec.~\ref{sec:cse}) that capture the structural information via maximizing the mutual information between a node and its surrounding subgraph.
Through GMML, SSA-Encoder (Sec.~\ref{sec:snfe}) models attribute correlations that play a pivotal role in predicting BMAs.
We train the four encoders simultaneously so that they can embed nodes into high-quality representations which are later used for detecting BMAs in the lightweight detection phase (Sec.~\ref{sec:put}).

\subsection{Graph Construction and Feature Extraction}
\label{sec:prepare}

\emph{BMA detection should not violate users' privacy}. Hence, \ours cannot access private information like users' chat content (text, video or speech) in \app. \emph{The data we used is chosen through a strict investigation process in order to protect users' privacy}.

We construct a user-user interaction graph where each node represents a \app account and each edge indicates that the connected nodes have some relations (e.g., they are in the same chat group, they are direct friends, or they have direct transactions).
The graph is homogeneous and multiple relations between two nodes, if they exist, are merged as one edge.
A more complex design like a heterogeneous graph~\cite{ZhongLAHFTH20} that distinguishes different node types and edge types can be used. 
However, using a homogeneous graph significantly reduces the number of nodes and edges, which is important for processing billion-scale accounts.
In the meantime, in our experiments, we observe that using a homogeneous graph already provides satisfactory detection performance.

For each node, we extract a $7$-dimensional binary vector where 0/1 in each dimension indicates the existence/non-existence of an attribute.  
Node attributes are designed by human experts and they may provide clues for detecting BMAs.

\subsection{Inductive Graph Neural Network Encoder}
\label{sec:gnn}

We design an \underline{I}nductive \underline{G}raph Neural Network Encoder (IG-Encoder) as the backbone in \ours. Note that the choice of the core GNN encoder is orthogonal to the design of SSS-Encoder, SS-Encoder and SSA-Encoder in \ours, and other GNNs can be used to replace IG-Encoder. However, in experiments (Sec.~\ref{sec:overall}), IG-Encoder shows much better performance than other GNNs for the BAD tasks.

Recall that each node $i$ is associated with a binary attribute vector $\mathbf{a}_i\in\mathbb{R}^{7}$. 
We use an \emph{attribute encoding mechanism} to encode node attributes.
For each attribute in $\mathbf{a}_i$, IG-Encoder maintains two $f$-dimensional feature vectors initialized from $\mathcal{N}(0,1)$ for its existence and non-existence. Thus, for each attribute vector $\mathbf{a}_i$, IG-Encoder has $7$ feature vectors and we concatenate them to construct the initial representation vector $\mathbf{h}^{(0)}_i\in\mathbb{R}^{7f}$ of the node $i$.

We adopt the idea of spatial GNNs~\cite{WuPCLZY20} for IG-Encoder. 
Firstly, information from neighboring nodes is aggregated:

\hspace{-15pt}
\begin{minipage}{.5\columnwidth}
	\begin{equation}
	\begin{aligned}
	\mathbf{x}_{\mathcal{N}_i}^{(t+1)} &=\text{mean}\big(\mathbf{h}_{j_1}^{(t)},\cdots,\mathbf{h}_{j_*}^{(t)}\big)\\
	\mathbf{y}_{\mathcal{N}_i}^{(t+1)} &=\text{max}\big(\mathbf{h}_{j_1}^{(t)},\cdots,\mathbf{h}_{j_*}^{(t)}\big)\\
	\mathbf{z}_{\mathcal{N}_i}^{(t+1)} &=\text{sum}\big(\mathbf{h}_{j_1}^{(t)},\cdots,\mathbf{h}_{j_*}^{(t)}\big)\nonumber 
	\end{aligned}
	\end{equation}
\end{minipage}
\begin{minipage}{.5\columnwidth}
	\begin{equation}
	\label{eq:agg}
	\small
	\begin{aligned}
	\mathbf{h}_{\mathcal{N}_i}^{(t+1)}&=\mathbf{x}_{\mathcal{N}_i}^{(t+1)} \oplus \mathbf{y}_{\mathcal{N}_i}^{(t+1)} \oplus \mathbf{z}_{\mathcal{N}_i}^{(t+1)}\\
	\mathbf{g}_{\mathcal{N}_i}^{(t+1)}&=\mathbf{W}_{\mathbf{g}}\mathbf{h}_{\mathcal{N}_i}^{(t+1)}+\mathbf{b}_{\mathbf{g}}
	\end{aligned}
	\end{equation}
\end{minipage}
where the superscript ``$(t)$'' indicates $t$-th iteration, $\oplus$ is the concatenation operation, $\mathcal{N}_i$ denotes the neighbor set of node $i$ and $j_{*}\in\mathcal{N}_i$. $mean(\cdot)$, $max(\cdot)$ and $sum(\cdot)$ are average pooling, max pooling and sum pooling, respectively. $\mathbf{W}_{\mathbf{g}}$ and $\mathbf{b}_{\mathbf{g}}\in\mathbb{R}^{f_1}$ are learnable parameters. We use two different $\mathbf{W}_{\mathbf{g}}$ ($\mathbf{W}_{\mathbf{g}}\in\mathbb{R}^{f_1\times f_2}$ for $\mathbf{h}_{\mathcal{N}_i}^{(1)}$ and $\mathbf{W}_{\mathbf{g}}\in\mathbb{R}^{f_1\times f_1}$ for $\mathbf{h}_{\mathcal{N}_i}^{(t)}$ ($t>1$)) and two different $\mathbf{b}_{\mathbf{g}}$ for different iterations in Eq.~\ref{eq:agg} since $\mathbf{h}_{\mathcal{N}_i}^{(1)}$ and $\mathbf{h}_{\mathcal{N}_i}^{(t)}$ ($t>1$) have different shapes (See Appendix~\ref{sec:hp}). 
We do not employ any attention mechanism that considers the different importance of neighbor nodes in neighbor aggregation since the number of nodes in WeChat graph is huge and attention mechanisms will incur additional overhead. 

Then, IG-Encoder adds a self-connection to each node so that the information from the original node attributes will not vanish during the message passing procedure:
\begin{equation}
\label{eq:self-con}
\mathbf{s}_i^{(t+1)} = \mathbf{W}_{\mathbf{s}}\mathbf{h}_i^{(0)}+\mathbf{b}_{\mathbf{s}},\,\,\,\,\mathbf{e}_i^{(t+1)} =\text{RELU}\big(\mathbf{g}_{\mathcal{N}_i}^{(t+1)} \oplus \mathbf{s}_i^{(t+1)}\big),
\end{equation}
where $\mathbf{W}_{\mathbf{s}}\in\mathbb{R}^{f_1\times f_2}$ and $\mathbf{b}_{\mathbf{s}}\in\mathbb{R}^{f_1}$ are learnable parameters, and $\text{RELU}(\cdot)$ is the Rectified Linear Unit. Although $\mathbf{W}_{\mathbf{s}}$ and $\mathbf{b}_{\mathbf{s}}$ are updated during training, IG-Encoder always uses the initial representation vector $\mathbf{h}^{(0)}_i$ in Eq.~\ref{eq:self-con} to retain the impact of the original node attributes that come from the knowledge of human experts. 

After that, IG-Encoder passes $\mathbf{e}_i$ to a feedforward neural network followed by an $L_2$ normalization:
\begin{equation}
\label{eq:l2}
\mathbf{q}_i^{(t+1)} = \text{RELU}\Big(\mathbf{W}_{\mathbf{q}}\mathbf{e}_i^{(t+1)}+\mathbf{b}_{\mathbf{q}}\Big),\,\,\,\,\,\mathbf{h}_i^{(t+1)} = \mathbf{q}_i^{(t+1)}  \Big/ \left\|\mathbf{q}_i^{(t+1)} \right\|,
\end{equation}
where $\mathbf{W}_{\mathbf{q}}\in\mathbb{R}^{f_1\times 2f_1}$ and $\mathbf{b}_{\mathbf{q}}\in\mathbb{R}^{f_1}$ are learnable parameters.

IG-Encoder stacks two of the above GNN layers (i.e., Eqs.~\ref{eq:agg},~\ref{eq:self-con} and \ref{eq:l2}) to extract the representation of node $i$. 
The output representation $\mathbf{h}_i^{(t+1)}$ is fed into a prediction component with a feedforward neural network to estimate the suspicious score of node $i$:
\begin{equation}
\label{eq:bg_predict}
p_i^{(t+1)}=\sigma\bigg({\mathbf{w}_{p_{(3)}}}^T\Big(\mathbf{W}_{p_{(2)}}\big(\mathbf{W}_{p_{(1)}} \cdot \mathbf{h}_i^{(t+1)} + \mathbf{b}_{p_{(1)}}\big)+ \mathbf{b}_{p_{(2)}}\Big)+b_{p_{(3)}}\bigg),
\end{equation}
where $\mathbf{w}_{p_{(3)}}\in\mathbb{R}^{f_1}$, $\mathbf{W}_{p_{(2)}}\in\mathbb{R}^{f_1\times 2f_1}$, $\mathbf{W}_{p_{(1)}}\in\mathbb{R}^{2f_1\times f_1}$, $\mathbf{b}_{p(1)}\in\mathbb{R}^{2f_1}$, $\mathbf{b}_{p(2)}\in\mathbb{R}^{f_1}$ and $b_{p(3)}$ are learnable parameters, and $\sigma(\cdot)$ is the sigmoid function. If $p_i$ is larger than a pre-defined threshold $\rho$, node $i$ will be labeled as a BMA.

IG-Encoder can be optimized with a standard binary cross entropy loss (denoted as $\mathcal{L}_{\text{IG}}$ in Fig.~\ref{fig:overview}) over \emph{labeled} nodes (in the training data) for the BAD task.
However, IG-Encoder does not learn node embeddings and use them for the BAD task. Instead, the global $7f$-dimensional feature
vectors for node attributes and the parameters of IG-Encoder are updated and
later used to produce representations for new nodes during detection. Hence,
graph representation learning based on IG-Encoder is indeed
\emph{inductive} (i.e, the trained model can be used for unseen
nodes), which is essential for the BAD task.
\ours, which uses IG-Encoder as its backbone for representation learning, is therefore also able to generate representations for unseen nodes.

\subsection{Contrastive Structural Encoders}
\label{sec:cse}

\begin{figure}[t]
\centering
\includegraphics[width=0.95\columnwidth]{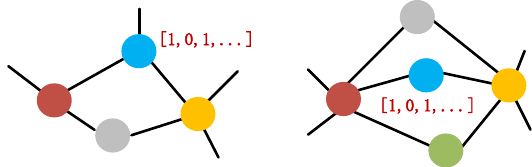}
\vspace{-10pt}
\caption{Two exemplifying surrounding structures: (1) Red node: BMA. (2) Blue node: Account for transactions. (3) Gray node: Account for delivering services. (4) Green node: Camouflage account. (5) Yellow node: Fraudster (buyer).}
\label{fig:subgraph}
\vspace{-10pt}
\end{figure}

Through manual check, we find that surrounding structures of nodes are useful in the BAD task.
Fig.~\ref{fig:subgraph} shows that, 
to avoid being easily detected, black market sellers manipulate several accounts for different purposes like camouflage, attracting fraudsters, transactions, delivering support services, etc.
However, these accounts are controlled by the same person/gang, and their surroundings often display similar characteristics. For example, for the purpose of transactions, suspect accounts with similar attributes (blue nodes in Fig.~\ref{fig:subgraph}) appear in the surrounding structures of various BMAs.

Although it is difficult to leverage manual rules of surrounding patterns, as discussed in P1 of Sec.~\ref{sec:intro}, the surrounding structures (i.e., subgraphs) indeed have rich clues for detecting BMAs. 
Hence, we believe it is beneficial to
maximize the correlation between a node and its subgraph and encode the subgraph as a context into node representation. 
Along this direction, we design contrastive structural encoders to maximize the
\emph{local} mutual information (MI) between the \emph{subgraph-level} summary $\mathbf{s}_i$ 
and the \emph{node-level} representation
$\mathbf{h}_i$ of the center node $i$ via graph contrastive
learning~\cite{abs-2103-00111}. 

We first use IG-Encoder to encode each node into node-level representation $\mathbf{h}$. 
Then, for building 
contrastive structural encoders, two issues remain: (1) How do we define the surrounding structure
of a node? (2) How to retrieve subgraph-level summary?

For the first issue, an elaborate way is to design representative subgraph
schemas. For instance, human experts can design subgraph schemas according to structural features (e.g., node degree or
graph density) and rule-based features (e.g, frequency of transactions).
Apparently, this is costly for a large graph. Moreover, as new data comes into
the graph and cybercriminals are trying different strategies to escape the
detection, schemas may soon become invalid.
Hence, we choose a simple yet effective design: use the complete
$k$-hop neighbors of the center node as its subgraph. There are other
sophisticated approaches to automatically retrieve subgraphs, 
e.g., use neighbors having largest personalized pagerank
values w.r.t. the center node~\cite{JiaoXZ0ZZ20}, 
or conduct random walks starting from the
center node~\cite{QiuCDZYDWT20}. But they also
introduce randomness: the subgraph for the
same center node may vary at each iteration, making preprocessing subgraphs
difficult. Differently, using complete $k$-hop neighbors can significantly reduce
the time to retrieve a subgraph since they are fixed and can be preprocessed.

For the second issue, given a $k$-hop subgraph, we deploy a \emph{readout} function $\mathcal{R}: \mathbb{R}^{n\times f_1}\rightarrow \mathbb{R}^{f_1}$, where $n$ is the number of nodes in the subgraph, to summarize $f_1$-dimensional subgraph-level representations. Here, we apply average pooling as the readout function: $\mathbf{s}_i=\frac{1}{\left|\mathcal{S}_i\right|}\sum_{j\in\mathcal{S}_i}\mathbf{h}_j$, 
where $\mathcal{S}_i$ is the set of all nodes in the $k$-hop subgraph of node $i$ (including $i$) and $\left|\mathcal{S}_i\right|$ indicates the number of nodes in $\mathcal{S}_i$.
Note that, from now, we omit the superscript ``(t)'' for simplicity.
Attention mechanisms can be employed in the readout function to distinguish different importance of nodes.  
But they typically require additional parameters and costly operations (e.g., the Softmax function when inputs contain many elements) that will incur high overhead on the large \app graph. Thus, we do not incorporate an attention mechanism in current \ours.

Then, we adopt a self-supervised MI objective for maximizing MI between the subgraph-level summary representation and the center node representation:
\begin{equation}
\label{eq:MI}
\mathcal{L}_{\text{MI}}=\sum_{i=1}^{N_{pos}}\left(\mathbb{E}_{pos}\left[\log\left(\mathcal{D}(\mathbf{h}_i,\mathbf{s}_i)\right)\right]+\mathbb{E}_{neg}\left[\log\left(1-\mathcal{D}(\mathbf{h}_{i'},\mathbf{s}_i)\right)\right]\right),
\end{equation}
where $N_{pos}$ is the number of nodes in the positive node set, $\mathbf{h}_{i'}$ is the representation of the negative node $i'$ w.r.t. to $i$ (we will discuss how to generate negative nodes later), and $\mathcal{D}(\mathbf{h}, \mathbf{s})$ is a discriminator that estimates the divergence and assigns probability to the node-subgraph representation pair $\left<\mathbf{h}, \mathbf{s}\right>$. 
We use a feedforward neural network as the discriminator but other designs can also be applied:
\begin{equation}
\label{eq:dis}
\mathcal{D}(\mathbf{h}_i,\mathbf{s}_i)=\sigma\Big({\mathbf{w}_{\mathcal{D}}^{(2)}}^T\big(\mathbf{W}_{\mathcal{D}}^{(1)} \cdot (\mathbf{h}_i \oplus \mathbf{s}_i) + \mathbf{b}_{\mathcal{D}}^{(1)}\big)+b_{\mathcal{D}}^{(2)}\Big),
\end{equation}
where $\mathbf{w}_{\mathcal{D}}^{(2)}\in\mathbb{R}^{f_1}$, $\mathbf{W}_{\mathcal{D}}^{(1)}\in\mathbb{R}^{f_1\times 2f_1}$, $\mathbf{b}_{\mathcal{D}}^{(1)}\in \mathbb{R}^{f_1}$ and $b_{\mathcal{D}}^{(2)}$ are learnable parameters. 
Eq.~\ref{eq:MI} is a noise-contrastive type objective with a standard binary cross-entropy loss. Optimizing Eq.~\ref{eq:MI} maximizes MI between $\mathbf{h}_i$ and $\mathbf{s}_i$ based on the Jensen-Shannon divergence between the joint distribution (positive samples) and the product of marginals (negative samples)~\cite{HjelmFLGBTB19,VelickovicFHLBH19}.

Depending on how positive and negative node-subgraph pairs are generated, two contrastive structural encoders are used in \ours:

\vspace{5pt}
\noindent\textbf{(1) SSS-Encoder}, short for \underline{S}elf-\underline{s}upervised Contrastive \underline{S}tructural Encoder, is designed for the complete self-supervised setting without any labels. SSS-Encoder treats each node $i$ and its subgraph $\mathcal{S}_i$ as a positive pair $\left<i, \mathcal{S}_i\right>$. For each $\left<i, \mathcal{S}_i\right>$, SSS-Encoder randomly selects another node $i'$, and use $i'$ and $\mathcal{S}_i$ as the negative pair $\left<i', \mathcal{S}_i\right>$. For SSS-Encoder, each node $i$ is a positive node. In the meantime, $i$ can be randomly selected as a negative node when SSS-Encoder encodes other nodes in the graph.

\vspace{5pt}
\noindent\textbf{(2) SS-Encoder}, short for \underline{S}upervised Contrastive \underline{S}tructural Encoder, is designed for the supervised setting, where a handful of labels are available. SS-Encoder uses each BMA $i$ and its subgraph $\mathcal{S}_i$ as a positive node-subgraph pair $\left<i, \mathcal{S}_i\right>$. For each $\left<i, \mathcal{S}_i\right>$, SS-Encoder randomly selects another node $i'$ from non-BMA set and use $i'$ and $\mathcal{S}_i$ as the negative pair $\left<i', \mathcal{S}_i\right>$. BMAs and non-BMAs are labeled by human experts.

\vspace{5pt}
For both encoders, negative
pairs are randomly regenerated at each iteration. 
IG-Encoder used in SSS-Encoder (SS-Encoder) for encoding nodes is denoted as IG-Encoder$_{\text{SSS}}$ (IG-Encoder$_{\text{SS}}$) in Fig.~\ref{fig:overview}. 
SSS-Encoder (SS-Encoder) is optimized using the loss function in Eq.~\ref{eq:MI} and we denote it as $\mathcal{L}_{\text{SSS}}$ ($\mathcal{L}_{\text{SS}}$) in Fig.~\ref{fig:overview}.
The output representation for node $i$ from SSS-Encoder (SS-Encoder) is indicated as $\mathbf{h}(\text{SSS})_i$ ($\mathbf{h}(\text{SS})_i$) in Fig.~\ref{fig:overview}.
We design two related but different contrastive
structural encoders so that human
experts are allowed to get involved in the detection. Although it is difficult
to manually label most nodes,
a small number of labeled nodes can be used as supervision in
SS-Encoder to improve representation learning. And we can observe from experiments (Sec.~\ref{sec:exp_encoder})
that using two encoders together can achieve better
performance than solely using either SSS-Encoder or SS-Encoder.

\subsection{Self-supervised Node Attribute Encoder}
\label{sec:snfe}

\begin{figure}[t]
\centering
\includegraphics[width=0.93\columnwidth]{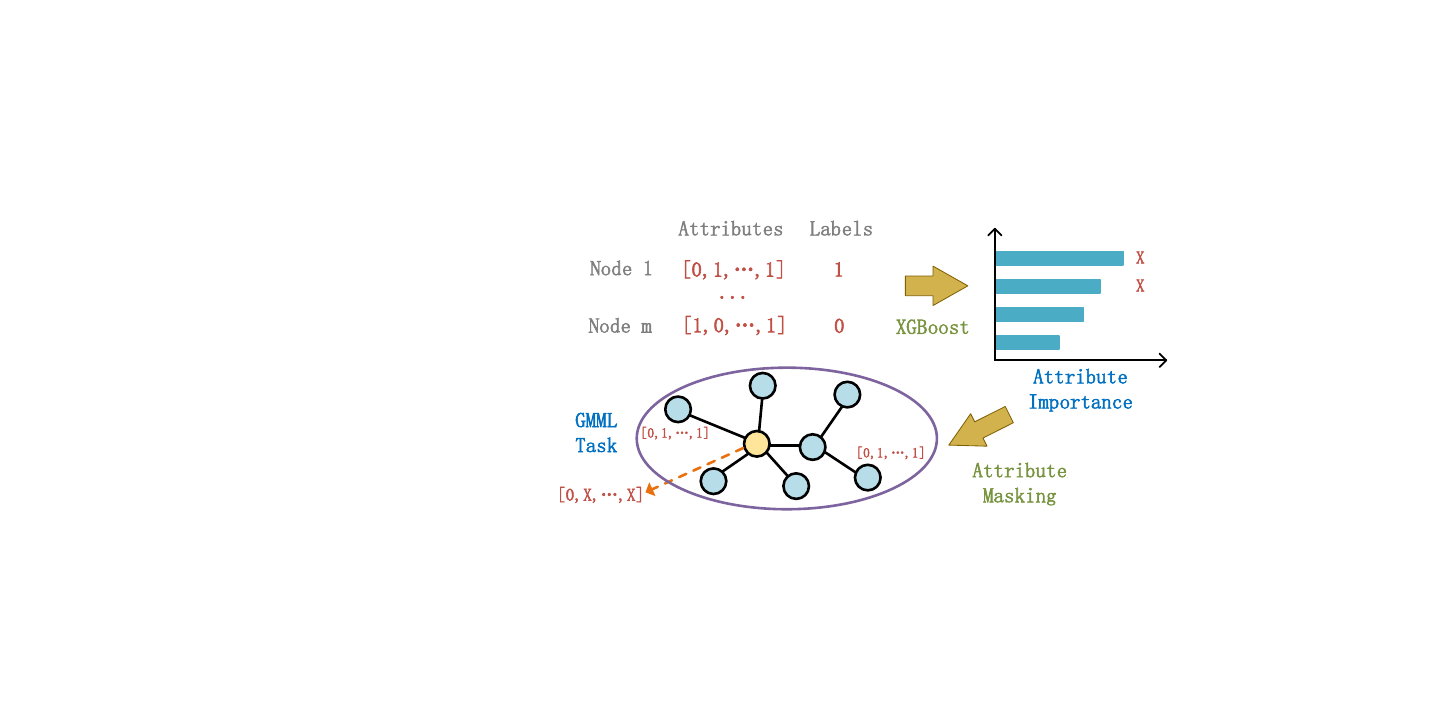}
\vspace{-10pt}
\caption{The workflow of SSA-Encoder.}
\label{fig:ssa}
\vspace{-15pt}
\end{figure}

We further design a \underline{S}elf-\underline{s}upervised Node
\underline{A}ttribute Encoder (SSA-Encoder). Although most labels are unavailable, we can use node attributes as the
attribute-level supervision to alleviate the reliance on the label-level
supervision. The intuition is that improving the prediction accuracy of some
node attributes (known), which are strongly correlated with node labels
(unknown), can help the attribute encoder capture attribute
correlations that benefit the downstream BAD task. Fig.~\ref{fig:ssa} depicts
the workflow of SSA-Encoder.
 
Specifically, we first adopt the non-GNN based method
XGBoost~\cite{ChenG16}, a gradient boosting algorithm that only relies on data attributes and shows promising
results in various prediction tasks, and train it on a small number of labeled
nodes to predict node labels. 
After training, XGBoost is
able to estimate the importance of each attribute to label prediction. Then, we pick two most crucial attributes to the prediction and treat them as \emph{pseudo labels} of each node. 
The reason for using more than one attribute as pseudo labels is that we find no attribute has a dominated
importance value. On the other hand, including more attributes as pseudo labels may
introduce noises 
since not all attributes have
high importance values. Hence, we choose the two most crucial attributes as pseudo labels to balance both sides.

After that, we formulate a \underline{G}raph-based \underline{M}ulti-label
prediction task with \underline{M}issing \underline{L}abels (GMML for short)
as a self-supervised learning task for SSA-Encoder: 
\begin{definition}[GMML]
Given partial attributes of the target node $i$, surrounding nodes in the
$k$-hop subgraph of $i$ and their complete node attributes, predict two
pseudo labels of node $i$.
\end{definition}

Multi-label learning with missing labels is well-studied for
multi-label image classification~\cite{LiuWST22}, but it receives less
attention in graph-relevant tasks. SSA-Encoder processes all nodes in the
graph in each iteration. For each node $i$, when SSA-Encoder predicts its
missing labels, we mask two dimensions of its attributes that correspond to pseudo labels as missing labels. When SSA-Encoder predicts missing labels of other nodes that have $k$-hop subgraphs containing $i$,
all node attributes of $i$ are known to SSA-Encoder.

SSA-Encoder uses an IG-Encoder (i.e., IG-Encoder$_\text{SSA}$) without the prediction component (Eq.~\ref{eq:bg_predict}) to encode a node $i$ into representation $\mathbf{\hat{h}}(\text{SSA})_i\in\mathbb{R}^{f_1}$ and then predicts pseudo labels of $i$:
\begin{equation}
\label{eq:ssa_pred}
\mathbf{r}_i=\sigma\Big(\mathbf{W}_{\mathbf{r}_{(2)}}\big(\mathbf{W}_{\mathbf{r}_{(1)}} \cdot \mathbf{\hat{h}}(\text{SSA})_i + \mathbf{b}_{\mathbf{r}_{(1)}}\big)+ \mathbf{b}_{\mathbf{r}_{(2)}}\Big),
\end{equation}
where $\mathbf{r}_i\in\mathbb{R}^2$ are the predicted probabilities of two pseudo labels for node $i$ being $1$. $\mathbf{W}_{\mathbf{r}_{(2)}}\in\mathbb{R}^{2\times\frac{f_1}{2}}$, $\mathbf{W}_{\mathbf{r}_{(1)}}\in\mathbb{R}^{\frac{f_1}{2}\times f_1}$ and $\mathbf{b}_{\mathbf{r}_{(1)}}\in\mathbb{R}^{\frac{f_1}{2}}$ and $\mathbf{b}_{\mathbf{r}_{(2)}}\in\mathbb{R}^2$ are learnable weights.

We utilize a multi-label prediction loss for optimizing SSA-Encoder: 
\begin{equation}
\mathcal{L}_{\text{SSA}}=-\frac{1}{2}\sum_{i=1}^{N}\sum_{j=1}^{2}\left[y_{ij}\log(r_{ij})+(1-y_{ij})\log(1-r_{ij})\right],
\end{equation}
where $N$ is the number of nodes in the graph, $y_{ij}$ indicates the $j$-th pseudo label of node $i$, and $r_{ij}$ is the predicted probability for the $j$-th pseudo label of node $i$ being 1.

Compared to GSSL-based methods~\cite{HuLGZLPL20} that randomly mask node features and then recover, SSA-Encoder predicts attributes that are crucial to node label prediction. Hence, it ``indirectly'' links attribute prediction to node label prediction via attribute importance.

\subsection{Putting All Together}
\label{sec:put}

Overall, detecting BMAs using \ours involves two phases:

\vspace{5pt}
\noindent\textbf{Pretraining Phase.} 
We pretrain SSS-Encoder, SS-Encoder and SSA-Encoder independently on the training data with limited labels to capture the intrinsic properties of \app graph from different aspects.
An IG-Encoder (i.e., IG-Encoder$_{\text{detect}}$) used later in detection is also trained over the limited training labels. All encoders can be optimized using gradient descent based optimization methods and we adopt Adam. Note that IG-Encoder$_{\text{SSS}}$, IG-Encoder$_{\text{SS}}$, IG-Encoder$_{\text{SSA}}$ and IG-Encoder$_{\text{detect}}$ do not share parameters.

\vspace{5pt}
\noindent\textbf{Detection Phase.} 
The four pretrained encoders are used for the \emph{lightweight} detection on the test data. 
For a target node $i$ (it may not exist in the training data), SSS-Encoder and SS-Encoder generate two node representations $\mathbf{h}(\text{SSS})_i\in\mathbb{R}^{f_1}$ and $\mathbf{h}(\text{SS})_i\in\mathbb{R}^{f_1}$.
We then adopt SSA-Encoder to predict pseudo labels for $i$. The predictions are used to replace the two attributes in $\mathbf{a}_i$ that correspond to pseudo labels to construct another attribute vector $\mathbf{a}'_i$ for $i$.
After that, we utilize the attribute encoding mechanism introduced in Sec.~\ref{sec:gnn} with feature vectors in IG-Encoder$_{\text{detect}}$ to encode $\mathbf{a}'_i$ into $\mathbf{h}(\text{SSA})_i\in\mathbb{R}^{7f}$.
Finally, the concatenation of $\mathbf{h}(\text{SSS})_i$ and $\mathbf{h}(\text{SS})_i$ and $\mathbf{h}(\text{SSA})_i$ is used as the initial representation for node $i$ and fed into IG-Encoder$_{\text{detect}}$ to estimate the suspicious score of node $i$:
\begin{equation}
\label{eq:init}
\mathbf{\bar{h}}_{i}^{(0)} =\mathbf{h}(\text{SSS})_i \oplus \mathbf{h}(\text{SS})_i \oplus \mathbf{h}(\text{SSA})_i,
\end{equation}
where $\mathbf{\bar{h}}_{i}^{(0)}\in\mathbb{R}^{2f_1+7f}$. 
In IG-Encoder$_{\text{detect}}$, shapes of $\mathbf{W}_g$ and $\mathbf{b}_g$ in Eq.~\ref{eq:agg} are modified according to the shape of $\mathbf{\bar{h}}^{(0)}_{i}$ (See Appendix~\ref{sec:hp}).

The above method that replaces true attributes with predicted attributes from SSA-Encoder outperforms using original attributes (see Sec.~\ref{sec:ssa_study}).
Recall that the \emph{direct} optimization objective of SSA-Encoder is to capture the correlation between partial attributes and pseudo labels.  
Since we choose two most crucial attributes to node label prediction as pseudo labels, predicting pseudo labels accurately can \emph{indirectly} increase the prediction accuracy of node labels (the ultimate goal of \ours). 
Nevertheless, for some nodes, the importance values of their pseudo labels do not dominate those of other attributes, i.e., the original attribute values for the two pseudo labels are not strongly connected to node labels and they are not ``good'' attributes for the BAD task.
For such cases, SSA-Encoder can give predictions that are not the ground-truth pseudo labels, but they follow captured attribute correlations which can lead SSA-Encoder to better benefit BMA detection.


\section{Discussion}
\label{sec:dis}

In this section, we provide some discussions about \ours:

\vspace{5pt}
\noindent\textbf{Model Size.} 
We provide the detailed analysis in Appendix~\ref{sec:model_size}.
The model size $\mathcal{S}_{\text{IG}}$ of IG-Encoder is in the magnitude of $f_1f_2$ where $f_1$ and $f_2$ are up to the order of tens (setting $f_1=32$ and $f_2=56$ in our experiments brings sufficiently good performance). 
Note that prevalent GNN-based methods involve at least one weight matrix in the neural network and their model size is at least $f_1f_2$. 
The model size $\mathcal{S}_{\text{\ours}}$ of the complete \ours, without considering XGBoost, is four to five times larger than $\mathcal{S}_{\text{IG}}$.
Nevertheless, $\mathcal{S}_{\text{\ours}}$ is still in the magnitude of $f_1f_2$. 
In practice, our BMA detection system is trained offline for the pretraining phase on small, million-scale data and then its detection phase is performed online on the large \app graph. The additional parameter cost of \ours, compared to using IG-Encoder only, is acceptable for the offline pretraining phase where training time and storage cost are not strictly limited. The online lightweight detection phase does not involve parameter update. Considering that \ours achieves much better performance than other methods (Sec.~\ref{sec:exp}), the model size of \ours is acceptable. In our experiments (Sec.~\ref{sec:online}), we also show that \ours can be deployed on billion-scale \app graph. 

\vspace{5pt}
\noindent\textbf{Relation to DGI and DCI.} The idea of \ours is closely related to Deep Graph Infomax (DGI)~\cite{VelickovicFHLBH19} and Deep Cluster Infomax (DCI)~\cite{Wang0GYL021}. 
DGI contrasts the whole graph with a node (i.e., \emph{global} MI) so that the global information can be embedded into node representation. 
When deploying on a large-scale graph like \app, summarizing the complete graph brings a global context with much noise. 
As discussed in Sec.~\ref{sec:cse}, the local subgraph provides rich clues and \ours contrasts the subgraph with the node in it.
In addition, \ours maximizes local MI instead of global MI, 
removing the obstacle to processing a billion-scale graph where summarizing the global graph-level representation has high cost. 
DCI can be viewed as a cluster-based DGI~\cite{Wang0GYL021} as it contrasts a cluster obtained from K-Means with the node in it. 
Although DCI reduces the context level from the complete graph to a cluster, a cluster in a large-scale graph is still too large to reveal BMA patterns.
Compared to DCI, \ours not only uses local surrounding structures in contrastive learning (SSS-Encoder and SS-Encoder) but also captures attribute-level correlations to benefit downstream BAD task (SSA-Encoder).

\vspace{5pt}
\noindent\textbf{Deployment challenge and design tradeoffs.} 
Due to the large volume of MMMA data in \app, the greatest deployment challenge is how to achieve good detection performance and avoid high overhead. 
\ours is currently retrained and updated on a daily basis to allow sufficient training time. 
Other design tradeoffs for reducing cost, which have been discussed in Sec.~\ref{sec:method}, include using a homogeneous graph instead of a heterogeneous graph (Sec.~\ref{sec:prepare}), avoiding using attention mechanisms (Secs.~\ref{sec:gnn} and~\ref{sec:cse}), using complete $k$-hop neighbors instead of sophisticated subgraph schemas (Sec.~\ref{sec:cse}) and using two pseudo labels (Sec.~\ref{sec:snfe}).


\section{Experiments}
\label{sec:exp}

\subsection{Experiment Setup}

\subsubsection{Data}
\label{sec:data}

In the \emph{offline test}, 
we use a million-scale dataset sampled from \app. 
The graph constructed on this dataset has 4 million nodes and 13.4 million edges.
100,000 nodes are manually labeled by human experts: half of them are BMAs and the remaining nodes are non-BMAs. The labels for other 3.9 million nodes are unknown.
Note that we also investigate the impact of the number of labels in our experiments (Sec.~\ref{sec:few_label}). 
We randomly divide labels into 80\%/20\% for training/test in the offline test. 
In the \emph{online test} on the billion-scale \app graph, the \ours pretrained on the training set of the offline data is directly deployed for detection.

\subsubsection{Baselines}
We compare \ours and its backbone IG-Encoder with the following state-of-the-art methods (details are provided in Appendix~\ref{sec:bs}): 
\begin{enumerate}[leftmargin=14pt,topsep=1pt,itemsep=0.3pt]
\item \textbf{Non-GNN classification methods}: XGBoost~\cite{ChenG16} and MLP. 

\item \textbf{GNN-based representation learning}: GCN~\cite{KipfW17}, GAT~\cite{VelickovicCCRLB18} and GeniePath~\cite{LiuCLZLSQ19}. 

\item \textbf{Self-supervised graph representation learning methods}: IG-Encoder$_{\text{MTL}}$~\cite{YouCWS20}, IG-Encoder$_{\text{M3S}}$~\cite{SunLZ20} and Deep Graph Infomax (DGI)~\cite{VelickovicFHLBH19}. IG-Encoder$_{\text{MTL}}$ enhances IG-Encoder using node clustering as additional self-supervised task and optimizes the encoder in a multi-task learning framework~\cite{YouCWS20}. IG-Encoder$_{\text{M3S}}$ leverages the self-training method~\cite{SunLZ20} to improve IG-Encoder.  

\item \textbf{Self-supervised graph anomaly detection method}: Deep Cluster Infomax (DCI)~\cite{Wang0GYL021}. Similar to IG-Encoder$_{\text{MTL}}$, DCI adopts node clustering as a self-supervised task. It also decouples representation learning phase and detection phase as \ours. 
\end{enumerate}

\vspace{5pt}
We modify MLP and GCN, which are originally transductive, with the attribute encoding mechanism used in IG-Encoder so that they can generalize to new data.  
In the reported results, we use abbreviations ``XGB'', ``IG'', ``SSS'', ``SS'' and ``SSA'' to indicate XGBoost, IG-Encoder, SSS-Encoder, SS-Encoder and SSA-Encoder, respectively. Other abbreviations are explained when they are used.

\subsubsection{Hyper-parameters and other settings}
Details are provided in Appendix~\ref{sec:hp}.
By default, we use $1$-hop subgraph in SSS-Encoder and SS-Encoder.
We test several values for the threshold $\rho$ used for judging whether the probability $p$ output in Eq.~\ref{eq:bg_predict} or other prediction layers in baselines indicates a BMA or not, and the threshold $r$ for predicting binary labels in SSA-Encoder according to $\mathbf{r}$ output in Eq.~\ref{eq:ssa_pred}. Finally, we set $\rho=0.5$ and $r=0.5$ which are sufficient for all methods to work well.
For a fair comparison, we set $f=64$ for the attribute encoding in all methods in the detection phase.
We use the Adam optimizer for all methods if applicable.
\emph{We tune hyper-parameters of all methods so that they achieve good results.}

\subsection{Offline Test} 
\label{sec:offline}

AUC, ACC (accuracy, i.e., the percentage of correctly predictions on test labels), KS (Kolmogorov-Smirnov statistic~\cite{massey1951kolmogorov}), Recall, Precision and F1-score are used for offline evaluation. 

\subsubsection{Overall Performance} 
\label{sec:overall}
Tab.~\ref{tab:results1} reports results of different approaches in the offline test and we can observe that:
\begin{enumerate}[leftmargin=14pt,topsep=1pt,itemsep=0.3pt]

\item IG-Encoder significantly outperforms other GNN-based methods GCN, GAT and GeniePath. This observation has supported our decision of using IG-Encoder as the backbone of \ours. \ours exceeds IG-Encoder by a large margin. Thus the backbone IG-Encoder is not the only reason for \ours's remarkable performance.

\item Compared to IG-Encoder, IG-Encoder$_{\text{MTL}}$ and IG-Encoder$_{\text{M3S}}$ generally have better results, showing the effectiveness of recently proposed self-supervised reinforcements~\cite{YouCWS20,SunLZ20} on the BAD task.

\item Compared to state-of-the-art self-supervised learning methods IG-Encoder$_{\text{MTL}}$, IG-Encoder$_{\text{M3S}}$, DGI and DCI, \ours consistently shows noticeably superior performance.

\end{enumerate}

\begin{table}[t]
\caption{Offline performance. Results of \ours and best baselines are shown in bold. Percentages indicate improvements of \ours over best baselines.}
\label{tab:results1}
\vspace{-12pt}
\scalebox{0.74}{
\begin{tabular}{|c|c|c|c|c|c|c|}
\hline
\textbf{Model}    & \textbf{AUC}                                                                 & \textbf{ACC}                                                                 & \textbf{KS}                                                                  & \textbf{Recall}                                                              & \textbf{Precision}                                                           & \textbf{F1-score}                                                            \\ \hline
XGB               & 0.6843                                                                       & 0.6605                                                                       & 0.3253                                                                       & 0.6342                                                                       & 0.6749                                                                       & 0.6540                                                                       \\ \hline
MLP               & 0.6851                                                                       & 0.6623                                                                       & 0.3260                                                                       & 0.6612                                                                       & 0.6678                                                                       & 0.6645                                                                       \\ \hline
GCN               & 0.7197                                                                       & 0.6607                                                                       & 0.3229                                                                       & 0.6272                                                                       & 0.6778                                                                       & 0.6516                                                                       \\ \hline
GAT               & 0.7514                                                                       & 0.6897                                                                       & 0.3802                                                                       & 0.6709                                                                       & 0.7022                                                                       & 0.6862                                                                       \\ \hline
GeniePath         & 0.7738                                                                       & 0.7014                                                                       & 0.4100                                                                       & 0.7123                                                                       & 0.7018                                                                       & 0.7070                                                                       \\ \hline
IG                & 0.7934                                                                       & 0.7223                                                                       & 0.4331                                                                       & 0.6665                                                                       & 0.7153                                                                       & 0.6901                                                                       \\ \hline
IG$_{\text{MTL}}$ & 0.7962                                                                       & 0.7241                                                                       & 0.4570                                                                       & 0.6550                                                                       & \textbf{0.7429}                                                              & 0.6962                                                                       \\ \hline
IG$_{\text{M3S}}$ & 0.7949                                                                       & \textbf{0.7243}                                                              & \textbf{0.4592}                                                              & \textbf{0.7154}                                                              & 0.7006                                                                       & 0.7080                                                                       \\ \hline
DGI               & 0.7897                                                                       & 0.6900                                                                       & 0.3818                                                                       & 0.6940                                                                       & 0.7098                                                                       & 0.7018                                                                       \\ \hline
DCI               & \textbf{0.8072}                                                              & 0.6987                                                                       & 0.3991                                                                       & 0.7138                                                                       & 0.7172                                                                       & \textbf{0.7155}                                                              \\ \hline
\ours             & \textbf{\begin{tabular}[c]{@{}c@{}}0.9402\\ $\uparrow$ 16.48\%\end{tabular}} & \textbf{\begin{tabular}[c]{@{}c@{}}0.8606\\ $\uparrow$ 18.82\%\end{tabular}} & \textbf{\begin{tabular}[c]{@{}c@{}}0.7263\\ $\uparrow$ 58.17\%\end{tabular}} & \textbf{\begin{tabular}[c]{@{}c@{}}0.8622\\ $\uparrow$ 20.52\%\end{tabular}} & \textbf{\begin{tabular}[c]{@{}c@{}}0.8622\\ $\uparrow$ 16.06\%\end{tabular}} & \textbf{\begin{tabular}[c]{@{}c@{}}0.8622\\ $\uparrow$ 20.50\%\end{tabular}} \\ \hline
\end{tabular}
}
\vspace{-5pt}
\end{table}

\begin{table}[t]
\caption{Offline Performance of different variations of \ours. Best performance is shown in bold.
}
\label{tab:results2}
\vspace{-12pt}
\scalebox{0.84}{
\begin{tabular}{|l|c|c|c|c|c|c|}
\hline
\multicolumn{1}{|c|}{\textbf{Model}} & \textbf{AUC}    & \textbf{ACC}    & \textbf{KS}     & \textbf{Recall} & \textbf{Precision} & \textbf{F1-score} \\ \hline
IG                                   & 0.7934          & 0.7223          & 0.4331          & 0.6665          & 0.7153             & 0.6901            \\ \hline
SSS                                  & 0.8754          & 0.7892          & 0.5790          & 0.7673          & 0.8064             & 0.7864            \\ \hline
SSS$_{\text{nn}}$                    & 0.8070          & 0.7330          & 0.4680          & 0.6945          & 0.7575             & 0.7246            \\ \hline
SS                                   & 0.8521          & 0.7695          & 0.5427          & 0.7254          & 0.8002             & 0.7610            \\ \hline
SS$_{\text{(2)}}$                    & 0.7893          & 0.7194          & 0.4419          & 0.6570          & 0.7562             & 0.7031            \\ \hline
SS$_{\text{sg}}$                     & 0.7884          & 0.7153          & 0.4421          & 0.6175          & 0.7739             & 0.6869            \\ \hline
SS$_{\text{ss}}$                     & 0.8402          & 0.7586          & 0.5230          & 0.6786          & 0.8068             & 0.7371            \\ \hline
SS$_{\text{nn}}$                     & 0.7900          & 0.7194          & 0.4426          & 0.6350          & 0.7700             & 0.6960            \\ \hline
SSA                                  & 0.8199          & 0.7412          & 0.4882          & 0.6421          & 0.8068             & 0.7151            \\ \hline
SGRL$_{\text{SA}}$                   & 0.9046          & 0.8143          & 0.6380          & 0.7833          & 0.8479             & 0.8143            \\ \hline
SGRL$_{\text{S}^2}$                  & 0.9104          & 0.8268          & 0.6559          & 0.8117          & 0.8403             & 0.8258            \\ \hline
\ours                                & \textbf{0.9402} & \textbf{0.8606} & \textbf{0.7263} & \textbf{0.8622} & \textbf{0.8622}    & \textbf{0.8622}   \\ \hline
\end{tabular}
}
\vspace{-5pt}
\end{table}

\subsubsection{Contributions of Each Encoder}
\label{sec:exp_encoder}
We show the performance of different variations of \ours in Tab.~\ref{tab:results2}: 
\begin{enumerate}[leftmargin=14pt,topsep=1pt,itemsep=0.3pt]
\item SSA in Tab.~\ref{tab:results2} indicates we uses SSA-Encoder and IG-Encoder$_{\text{detect}}$ together for detection. We can see that all encoders (SSS, SS, SSA) in \ours exceed original IG-Encoder. But using any single encoder does not perform well as using the complete \ours. 

\item In Tab.~\ref{tab:results2}, SGRL$_{\text{S}^2}$ denotes that the output representations from original attribute encoding, SSS-Encoder and SS-Encoder are concatenated as node representations.
We can find that using SSS-Encoder and SS-Encoder together (i.e., SGRL$_{\text{S}^2}$) yields better results than only using either SSS-Encoder or SS-Encoder.

\item Incorporating SSA-Encoder into $\text{SGRL}_{\text{S}^2}$ (i.e., \ours in Tab.~\ref{tab:results2}) achieves better performance than solely using $\text{SGRL}_{\text{S}^2}$.
\end{enumerate}

\vspace{5pt}
Based on above observations, we can conclude that each encoder indeed contributes to the performance of \ours.

\begin{figure}[t]
\centering
\begin{minipage}{.44\columnwidth}
  \centering
  \includegraphics[width=1\linewidth]{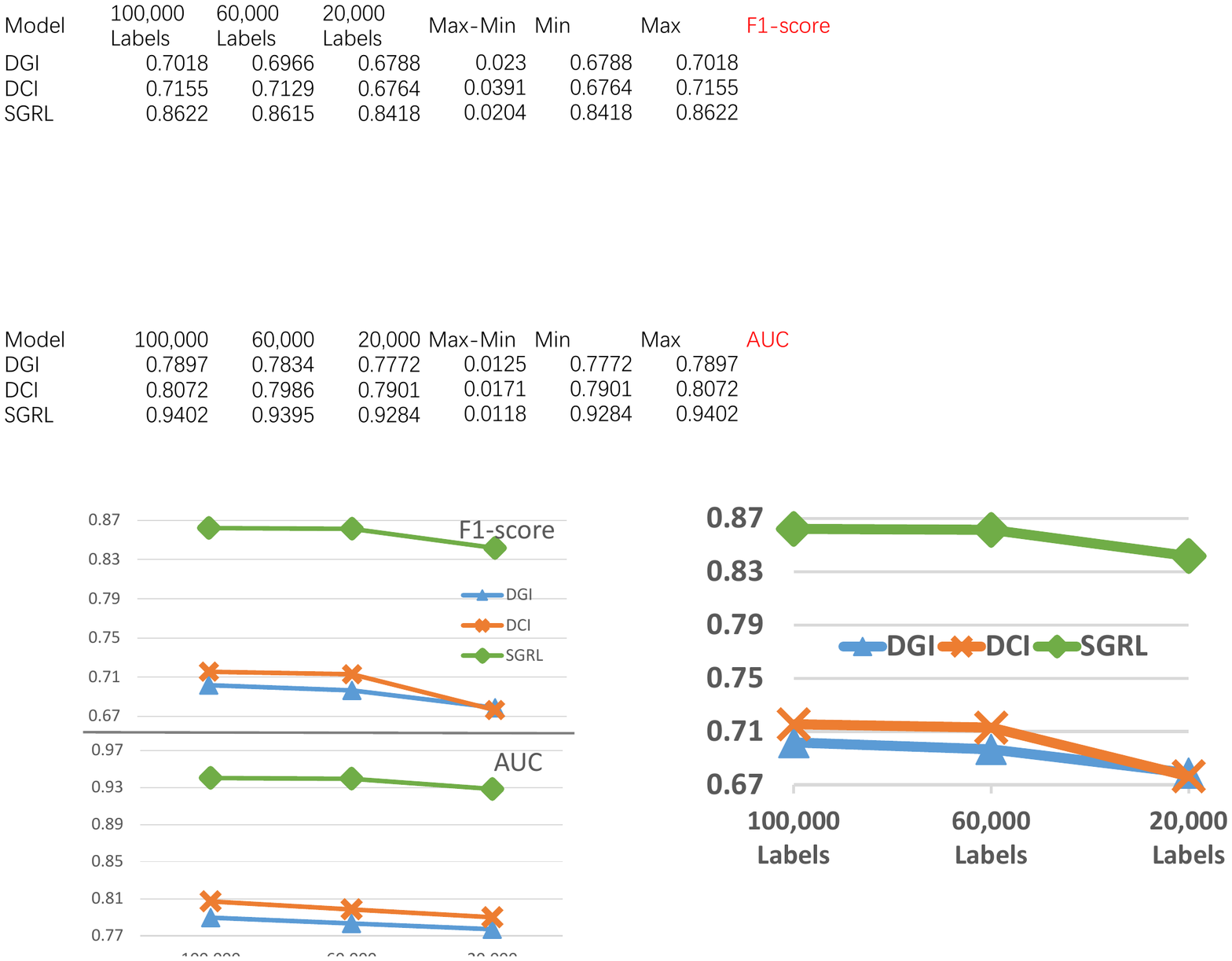}
  \vspace{-15pt}
  \captionof{figure}{F1-score of DGI, DCI and \ours when reducing number of labels.}
  \vspace{-10pt}
  \label{fig:label_test}
\end{minipage}%
\hspace{10pt}
\begin{minipage}{.47\columnwidth}
  \centering
  \includegraphics[width=1\linewidth]{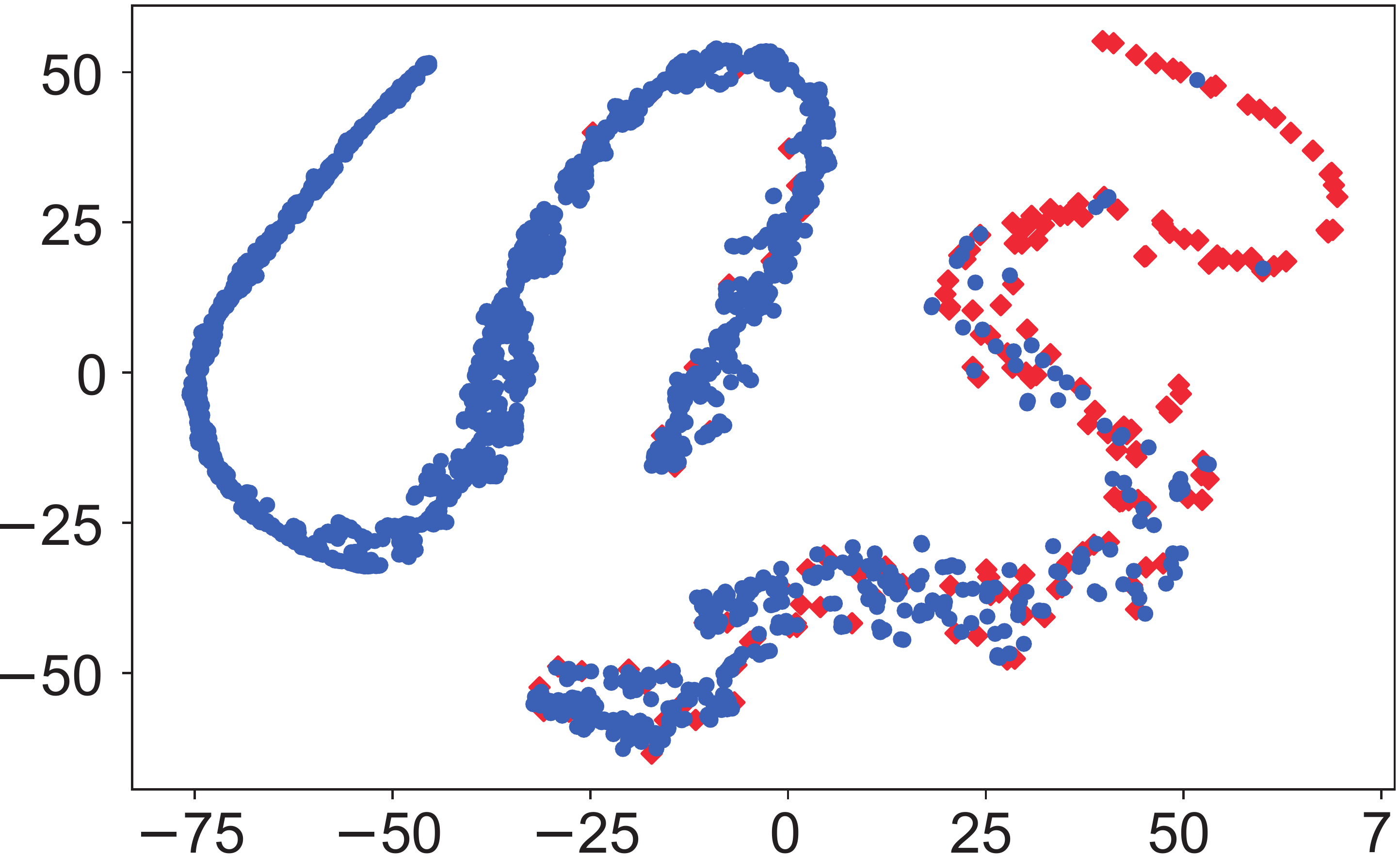}
  \vspace{-15pt}
  \captionof{figure}{Representation visualization: (1) Red: BMAs. (2) Blue: non-BMAs.}
  \vspace{-10pt}
  \label{fig:vis}
\end{minipage}
\end{figure}

\subsubsection{Results Using Fewer or No Labels} 
\label{sec:few_label}

Labels in the BAD task is scarce and the robustness of the detection methods, when facing fewer or no labels, is crucial to their practicability.
We randomly mask labels in the offline dataset and derive two new datasets with same numbers of nodes and edges as the original dataset. One retains 60,000 node labels and the other retains 20,000 node labels. Labels are randomly divided into 80\%/20\% for training/test as the original dataset with 100,000 node labels.
Fig.~\ref{fig:label_test} reports F1-score of DGI, DCI and \ours on three datasets.
We can find that \ours drops \emph{insignificantly} as we reduce the number of labels and it consistently outperforms DGI and DCI.
We also test \ours when no labels are available. For such cases, \ours degrades to SGRL$_{\text{SA}}$ (i.e., the outputs from SSA-Encoder and SSS-Encoder are concatenated as node representations) reported in Tab.~\ref{tab:results2}. 
We can see that SGRL$_{\text{SA}}$ performs well without labels and it is even better than those baselines reported in Tab.~\ref{tab:results1} that uses labels.
In summary, regardless of available labels, \ours can consistently provide accurate predictions.

\subsubsection{Visualization of Representations}
\label{sec:vis}

We adopt t-SNE~\cite{MaatenH08} to visualize node representations generated by \ours for the test set. Fig.~\ref{fig:vis} depicts the result: most BMAs are distributed in the right while other non-BMAs are located in the left and middle. Representations of different accounts  have a clear distinction, showing that \ours produces high-quality node representations for the BAD task.

\subsubsection{Choice of MI} 
\label{sec:mi}

We choose to maximize the local MI between a node and its subgraph in SSS-Encoder and SS-Encoder. There are other options for the MI.
In the following, ``$\leftrightarrow$'' indicates maximizing MI between the connected two objects. 
In Tab.~\ref{tab:results2}, we show performance of other possible choices for MI:
\begin{itemize}[leftmargin=10pt,topsep=1pt,itemsep=0.3pt]
	\item \textbf{subgraph$\leftrightarrow$global graph (SS$_{\text{sg}}$)}: We use the subgraph representation for a node $i$ and the global graph as a positive pair, and the subgraph representation for a randomly chosen node $i'$ with opposite label to $i$ and the global graph as the negative pair in Eq.~\ref{eq:MI} for SS-Encoder. Apparently, the global graph contains too much noise and it is hard to tell whether the MI between a positive subgraph and the global graph should be maximized. Therefore, this option shows worst performance for SS-Encoder. 

	\item \textbf{subgraph$\leftrightarrow$subgraph (SS${_\text{ss}}$)}: We maximize MI between subgraphs of nodes with same labels in SS-Encoder. According to Tab.~\ref{tab:results2}, this alternative design ranks the second among all SS-Encoder variations, showing the importance of subgraphs in the BAD task. For SSS-Encoder, we do not know a way to define the positive subgraph pairs since node labels are unknown. Hence, we do not experiment SSS${_\text{ss}}$.

	\item \textbf{node$\leftrightarrow$node (SSS${_\text{nn}}$ and SS${_\text{nn}}$)}: We maximize MI between two randomly picked nodes (SSS-Encoder) or nodes with same labels (SS-Encoder). These variations show much worse performance compared to default SSS-Encoder and SS-Encoder. This observation, again, shows the importance of subgraphs. 

\end{itemize}

\vspace{5pt}
As a summary, our design that maximizes the local MI between a node and its subgraph is most appropriate for \ours.

\subsubsection{Choice of $k$ for Subgraph} 
\label{sec:k}

By default, we use $1$-hop subgraphs ($k=1$). We also report the result of SS-Encoder for $k=2$, i.e., SS$_{(2)}$ in Tab.~\ref{tab:results2}: it noticeably degrades the performance of SS-Encoder. A possible reason is that larger subgraphs introduce noise and first-order neighbors are more important in subgraphs. Since we do not use any attention mechanism to distinguish different importance of nodes in subgraphs, using larger subgraphs degrades the performance. 
Besides, larger $k$ means longer time and larger memory cost for the pooling operation over subgraphs. When setting $k=2$,  
the cost is unaffordable and that is why we do not report SSS$_{(2)}$.

\subsubsection{Inputs to Pretrained SSA-Encoder} 
\label{sec:ssa_study}

By comparing SSA and IG (it is fed with original node attributes) in Tab.~\ref{tab:results2}, we can see that SSA-Encoder show better performance. As mentioned in Sec.~\ref{sec:put}, SSA-Encoder can predict pseudo labels that may not be the same as the original attribute values of pseudo labels but follow the captured attribute correlations which help SSA predict node labels better.

\subsection{Online Test} 
\label{sec:online}

In the online test, we deploy IG-Encoder and \ours that are pretrained on the million-scale ``small'' data with 100,000 labels used in the offline test to detect BMAs over the large, billion-scale \app graph. 
In other words, we do not perform the pretraining phase. 

In the online environment, we do not have test labels beforehand.
In practice, BMAs and non-BMAs are severely imbalanced, and normal detection methods will label most accounts as non-BMAs.
Thus, human experts do not check the large number of predicted non-BMAs as such manual checks are highly costly, and non-BMAs do not trigger warming in the detection system anyway.
However, human experts will check all the predicted BMAs before they are banned by the system to avoid disturbing normal \app users.
Considering the above online scenario where predicted BMAs are more important, we define an online evaluation measure called \emph{Detection Success Rate (DSR)} as the evaluation measure. DSR is the percentage that predicted BMAs are BMAs in reality.

IG-Encoder and \ours were deployed to detect BMAs over the \app graph once every day within a month. After the test period, we summarize each method's one-month result and remove duplicated predicted BMAs (an account can be labeled as a BMA on different days). Then, we measure DSR on the summarized result for each method.
The resulting DSR is 0.7495 and 0.8040 for IG-Encoder and \ours, respectively.
We can see that both methods achieve satisfactory performance in the online test. 
\ours further exceeds IG-Encoder by 7.27\%, showing its superiority.
Due to security reasons, we are not allowed to disclose the exact number of identified BMAs and detailed statistics of the \app graph.  
But we can roughly report that both methods correctly discover several hundred thousand BMAs over the large \app graph.
This is an encouraging result as a few human experts may not correctly identify ``unique'' BMAs in the same magnitude within a month. 
Moreover, most of the identified BMAs are not even included in the ``small'' training set used in the offline pretraining phase, meaning that our system has a strong generalization ability and it can adapt properly to new, unseen data.
In conclusion, the online test demonstrates that our system can effectively detect BMAs in \app.


\section{Related Work}
\label{sec:related}

\vspace{5pt}
\noindent\textbf{Graph Self-Supervised Learning (GSSL).}
Previous works on graph representation learning mostly focus on (semi-)supervised learning, resulting in heavy label reliance and poor generalization~\cite{abs-2103-00111}.
Recently, GSSL~\cite{YouCWS20,SunLZ20,JiaoXZ0ZZ20}, which extracts useful information from pretext tasks that do not rely on manual labels, has attracted great interest.
Various pretext tasks are explored to alleviate label reliance and strengthen model generalization. 
Examples include masking features and predicting~\cite{YouCWS20,HuLGZLPL20}, contrasting graph instances~\cite{VelickovicFHLBH19,YouCSCWS20,QiuCDZYDWT20}, node clustering~\cite{YouCWS20,SunLZ20}, and graph partitioning~\cite{YouCWS20}.
The above works have inspired the design of \ours.

\vspace{5pt}
\noindent\textbf{Graph-based Anomaly Detection (GBAD).}
GBAD has many applications~\cite{AkogluTK15} and various
techniques have been used to improve GBAD: adopt Heterogeneous Information Networks~\cite{ZhongLAHFTH20,XuSSACC21,XuSHQLL22}, use GSSL to decouple learning and detection~\cite{Wang0GYL021}, 
capture the temporal information~\cite{YuCAZCW18,ZhengLLLG19}, to name a few. Compared to these methods, \ours encodes both local structure and attribute-level correlations into node representations and scales well to large-scale graphs.

\vspace{5pt}
\noindent\textbf{Black Market Account Detection in Social Networks.} 
Lee et al~\cite{LeeWKK18,WooKKKK11} study real money trading detection in online games. 
But their methods are rule-based and inflexible. 
Supervised classification~\cite{DuttaCJ018} 
and Weighted Generalized Canonical Correlation Analysis~\cite{AroraDJCC20} are adopted to identify black market activities in Twitter. Compared to \ours, these methods heavily rely on annotations that are almost unavailable in the BAD task in \app.


\section{Conclusion}
\label{sec:con}

In this paper, we illustrate our BMA detection system \ours used in \app. 
\ours is designed based on GNN and GSSL. 
When detecting BMAs in \app, \ours shows not only promising detection results but also a great generalization ability. 
Both offline and online experiments demonstrate that \ours is able to detect BMAs over the large MMMA data in \app and block their supports for frauds.
In the future, we plan to explore how to reduce the cost of advanced designs mentioned in Sec.~\ref{sec:dis} that are not used in current \ours, and further improve the detection performance of \ours.

\begin{acks}
This work was supported by the National Natural Science Foundation of China (No. 62002303, 42171456), the Natural Science Foundation of Fujian Province of China (No. 2020J05001), and 2021 Tencent WeChat Rhino-Bird Focused Research Program.
\end{acks}

\bibliographystyle{ACM-Reference-Format}
\balance
\bibliography{main}


\appendix

\section{Hyper-parameters and Other Settings Used in Experiments}
\label{sec:hp}

We stop optimizing baselines, IG-Encoder and its variations (IG-Encoder$_{\text{MTL}}$, IG-Encoder$_{\text{M3S}}$) when the AUC value starts to decrease.
For \ours and its variations except SSA-Encoder, we concatenate the node representation $\mathbf{h}_i$ output by each method after each iteration and the node attribute vector $\mathbf{a}_i$ for node $i$. Then, we feed the result into XGBoost (it uses the same hyper-parameters as the baseline XGBoost described below) for node label predictions and evaluate the AUC value. When the AUC value starts to decrease, we stop training.
For SSA-Encoder, we feed the replaced attribute vector $\mathbf{a}_i$ into XGBoost for node label predictions and evaluate the AUC value. When the AUC value starts to decrease, we stop training SSA-Encoder.

We test several values for the threshold $\rho$ used for judging whether the probability $p_i$ output in Eq.~\ref{eq:bg_predict} or other prediction layers in baselines indicates a BMA or not, and the threshold $r$ for predicting binary labels in SSA-Encoder according to $\mathbf{r}_i$ output in Eq.~\ref{eq:ssa_pred}. Finally, we set $\rho=0.5$ and $r=0.5$ which are sufficient for all methods to work well.
For a fair comparison, we set $f=64$ for the attribute encoding in all baselines in the detection phase.

We tune hyper-parameters of all methods so that they achieve good results.
The hyper-parameter settings for different methods are listed as follows:
\begin{itemize}[leftmargin=10pt,topsep=1pt,itemsep=0.3pt]
     \item\textbf{XGBoost}: We use its official APIs\footnote{\url{https://xgboost.readthedocs.io}} and try to tune its hyper-parameters. The final settings used are: $\text{colsample\_bytree}=0.8$, $\text{learning\_rate}=0.2$, $\text{max\_depth}=4$, $\text{subsample}=0.9$, $\text{n\_estimators}=300$, $\text{early\_stopping\_rounds}=50$, and $\text{eval\_metric}$\\=``auc''.

     \item\textbf{MLP}: We first adopt the same attribute encoding mechanism as IG-Encoder to construct the initial representation vector $\mathbf{h}^{(0)}\in\mathbb{R}^{7f}$ ($f=64$). Then, we use 128, 64 and 32 neurons in the first hidden layer, the second hidden layer and the third hidden layer of the feedfoward neural network, respectively.

     \item\textbf{GCN}: We use ``dgl.nn.GraphConv'' in the DGL library\footnote{\url{https://www.dgl.ai}} to implement the baseline GCN. We first adopt the same attribute encoding mechanism as IG-Encoder to construct the initial representation vector $\mathbf{h}^{(0)}\in\mathbb{R}^{7f}$ ($f=64$). Then, we adopt two GCN layers to encode each node. The first GCN layer has an input feature size of $576$ and an output feature size of $128$. The second GCN layer has an input feature size of $128$ and an output feature size of $128$. The output representations are fed into the prediction component (Eq.~\ref{eq:bg_predict}). The prediction component has 64 and 32 neurons in the first hidden layer and the second hidden layer in Eq.~\ref{eq:bg_predict}, respectively.

     \item\textbf{GAT}: 
     We use ``dgl.nn.pytorch.conv.GATConv'' in the DGL library to implement the baseline GAT. 
     We adopt two GAT layers to encode each node. The first GAT layer has an input feature size of $576$, an output feature size of $128$ and 3 heads. The second GAT layer has an input feature size of $384$, an output feature size of $128$ and 3 heads. The output representations are fed into a feedforward neural network which is similar to Eq.~\ref{eq:bg_predict}. It has 64 and 32 neurons in the first hidden layer and the second hidden layer, respectively.

     \item\textbf{GeniePath}: We conduct grid search to find good hyper-parameters and finally we use 2 layers and 3 heads for GeniePath.

     \item\textbf{DGI}: We try different numbers of layers and find using 2 layers achieve good results.

     \item\textbf{DCI}: We conduct grid search to find best value for the number of clusters. After tuning, we set the number of clusters to be 25 and use 2 layers in DCI.

     \item\textbf{IG-Encoder, IG-Encoder$_{\text{MTL}}$ and IG-Encoder$_{\text{M3S}}$}: We set $f=64$ and map each node attribute to two $64$-dimensional vectors in attribute encoding. We set $f_1=128$ and $f_2=448$ ($64\times7$) for parameters. 
     We use 64 and 32 neurons in the first hidden layer and the second hidden layer in Eq.~\ref{eq:bg_predict}, respectively.

     \item\textbf{\ours and its variations}: 
     \begin{itemize}[leftmargin=10pt,topsep=1pt,itemsep=0.3pt]
          \item\textbf{Representation Learning Phase:} Using a relatively larger $f$ (e.g., use $f=64$ as IG-Encoder) generally improves the result. However, to reduce the cost of memory when handling a large graph, we set $f=8$ in the representation learning phase of SS-Encoder, SSS-Encoder and SSA-Encoder, and map each node attribute to two $8$-dimensional vectors in attribute encoding. We set $f_1=32$ and $f_2=56$ ($8\times7$) accordingly. Note that, to achieve a fair comparison between \ours and baselines ($f$ is set to $64$ when using baselines for detection), we set $f=64$ in the detection phase of \ours, i.e., set $f=64$ for IG-Encoder$_{\text{detect}}$ (see below). 
          
          \item\textbf{Detection Phase:} As described in Eq.~\ref{eq:init} of Sec.~\ref{sec:put}, in the detection phase of \ours, the initial concatenated node presentation $\mathbf{\bar{h}}_{i}^{(0)} =\mathbf{h}(\text{SSS})_i \oplus \mathbf{h}(\text{SS})_i \oplus \mathbf{h}(\text{SSA})_i$ for node $i$ has a dimensionality of $2f_1+7f$ where $f_1=32$ and $f=64$. 
     \end{itemize}

\end{itemize}

\section{Analysis of model size of \ours}
\label{sec:model_size}

\ours involves four graph-based encoders: SSS-Encoder, SS-Encoder, SSA-Encoder and IG-Encoder$_{detect}$, and they are all built on the top of IG-Encoder illustrated in Sec.~\ref{sec:gnn}.

\subsection{Model size of IG-Encoder}
The model parameters of one GNN layer in an IG-Encoder are $\mathbf{W}_{\mathbf{g}}\in\mathbb{R}^{f_1\times f_2}$ ($t=1$), $\mathbf{W}_{\mathbf{g}}\in\mathbb{R}^{f_1\times f_1}$ ($t>1$) and two $\mathbf{b}_{\mathbf{g}}\in\mathbb{R}^{f_1}$ in Eq.~\ref{eq:agg}, $\mathbf{W}_{\mathbf{s}}\in\mathbb{R}^{f_1\times f_2}$ and $\mathbf{b}_{\mathbf{s}}\in\mathbb{R}^{f_1}$ in Eq.~\ref{eq:self-con}, and $\mathbf{W}_{\mathbf{q}}\in\mathbb{R}^{f_1\times 2f_1}$ and $\mathbf{b}_{\mathbf{q}}\in\mathbb{R}^{f_1}$ in Eq.~\ref{eq:l2}. Hence, the parameter size of one GNN layer in the IG-Encoder is $(3f_1+2f_2+4)f_1$.

The prediction layer in Eq.~\ref{eq:bg_predict} has parameters 
$\mathbf{w}_{p_{(3)}}\in\mathbb{R}^{f_1}$, $\mathbf{W}_{p_{(2)}}\in\mathbb{R}^{f_1\times2f_2}$, $\mathbf{W}_{p_{(1)}}\in\mathbb{R}^{2f_1\times f_1}$, $\mathbf{b}_{p_{(1)}}\in\mathbb{R}^{2f_1}$, $\mathbf{b}_{p_{(2)}}\in\mathbb{R}^{f_1}$ and $b_{p_{(3)}}$, and its parameter size is $(4f_1+4)f_1+1$.

Denote the number of GNN layers as $l$ ($l\geq 2$). The total model size of IG-Encoder is $\mathcal{S}_{\text{IG}}=(3l\cdot f_1+2l\cdot f_2+4f_1+4l+4)f_1+1$ and it is in the magnitude of $f_1f_2$.

\subsection{Model size of SSS-Encoder and SS-Encoder}
The additional parameters of SSS-Encoder and SS-Encoder, compared to IG-Encoder, are the weight matrices ($\mathbf{w}_{\mathcal{D}}^{(2)}\in\mathbb{R}^{f_1}$ and $\mathbf{W}_{\mathcal{D}}^{(1)}\in\mathbb{R}^{f_1\times 2f_1}$) and bias terms ($\mathbf{b}_{\mathcal{D}}^{(1)}\in\mathbb{R}^{f_1}$ and $b_{\mathcal{D}}^{(2)}$) in Eq.~\ref{eq:dis}. Thus the model size of SSS-Encoder/SS-Encoder can be estimated as $\mathcal{S}_{\text{IG}}+2f_1f_1+2f_1+1$.

\subsection{Model size of SSA-Encoder}

SSA-Encoder does not include the prediction layer with a parameter size of $(4f_1+4)f_1+1$ in IG-Encoder (Eq.~\ref{eq:bg_predict}). Instead, it uses another prediction layer (Eq.~\ref{eq:ssa_pred}) to predict the two pseudo labels and the additional parameters are $\mathbf{W}_{\mathbf{r}_{(2)}}\in\mathbb{R}^{2\times\frac{f_1}{2}}$, $\mathbf{W}_{\mathbf{r}_{(1)}}\in\mathbb{R}^{\frac{f_1}{2}\times f_1}$, $\mathbf{b}_{\mathbf{r}_{(1)}}\in\mathbb{R}^{\frac{f_1}{2}}$ and $\mathbf{b}_{\mathbf{r}_{(2)}}\in\mathbb{R}^{2}$. The parameter size of Eq.~\ref{eq:ssa_pred} is $(\frac{3}{2}+\frac{f_1}{2})f_1+2$. Therefore, the model size of SSA-Encoder without considering XGBoost is $\mathcal{S}_{\text{IG}}-\frac{7}{2}f_1f_1-\frac{5}{2}f_1+1$. Note that it is difficult to estimate the exact model size of XGBosst as it depends on the implementation. The official APIs provide memory efficient XGBoost that can be used as long as the memory can accommodate the input data\footnote{\url{https://xgboost.readthedocs.io/en/stable/faq.html\#i-have-a-big-dataset}}.

\subsection{Model size of IG-Encoder$_{\text{detect}}$}

IG-Encoder$_{\text{detect}}$ uses the identical design as IG-Encoder. Hence, the model size of IG-Encoder$_{\text{detect}}$ is $\mathcal{S}_{\text{IG}}$.

\subsection{Model size of the complete \ours}
Sum up model sizes of SSS-Encoder, SS-Encoder, SSA-Encoder and IG-Encoder$_{\text{detect}}$, the model size of \ours without counting XGBoost is $\mathcal{S}_{\text{\ours}}=4\mathcal{S}_{\text{IG}}+\frac{1}{2}f_1f_1+\frac{3}{2}f_1+3$, i.e., $(12l\cdot f_1+8l\cdot f_2+\frac{33}{2}f_1+16l+\frac{35}{2})f_1+7$ where $l\geq2$. Although the model size of \ours is four to five times larger than the model size of IG-Encoder, it is still in the magnitude of $f_1f_2$.

\section{Descriptions of Baselines}
\label{sec:bs}

We modify MLP and GCN, which are originally transductive, with the attribute encoding mechanism used in IG-Encoder so that they can generalize to new data. 
The attribute encoding mechanism encodes each node $i$ into its initial representation $\mathbf{h}_i\in\mathbb{R}^{7f}$.
We retain parameters of graph encoders and the attribute feature vectors instead of node embeddings so that trained graph encoders can produce representations for unseen nodes.

The descriptions of baselines are provided as follows: 
\begin{enumerate}[leftmargin=14pt,topsep=3pt,itemsep=0.3pt]
	\item \textbf{XGBoost}~\cite{ChenG16} is a gradient boosting algorithm that shows promising results in various prediction tasks. It only relies on data attributes for prediction.

	\item \textbf{MLP} is a feedforward neural network with three hidden layers to predict the suspicious score of a node $i$. We adopt the Rectified Linear Unit in each hidden layer and the final output is passed through the sigmoid function.

	\item \textbf{GCN} uses Graph Convolutional Network~\cite{KipfW17} as the representation learning component to model node properties and graph structures. The output node representations are fed into the same prediction layer (Eq.~\ref{eq:bg_predict}) as IG-Encoder to estimate the suspicious score of a node $i$. 

	\item \textbf{GAT} is similar to GCN except that its representation learning component employs Graph Attention Network (GAT)~\cite{VelickovicCCRLB18}. 

     \item \textbf{GeniePath}\footnote{\url{https://github.com/shawnwang-tech/GeniePath-pytorch}}~\cite{LiuCLZLSQ19} uses path layers with adaptive breadth and depth functions to guide the receptive paths in graph representation learning. It works in both transductive and inductive settings. The output is fed into a single-layer feedforward network to estimate the suspicious score of a node $i$. 

     \item \textbf{DGI}\footnote{\url{https://github.com/PetarV-/DGI}}~\cite{VelickovicFHLBH19} is a self-supervised representation learning method that maximizes the mutual information between the node representation and the high-level summary of the global graph. To generate negative samples, it corrupts subgraphs by row-wise shuffling the feature matrices within subgraphs. The generated node representation is fed into a simple linear (logistic regression) classifier to predict the label of the node.

     \item \textbf{IG-Encoder$_{\mathbf{MTL}}$} is a multi-task learning method that trains IG-Encoder for both the BAD task and an auxiliary, self-supervised task~\cite{YouCWS20}. Due to the page limit, we report the result of using node clustering task as the auxiliary task since it shows better performance than other tasks proposed by You et al.~\cite{YouCWS20} in our experiments. We use the siamese network for IG-Encoder$_{\text{MTL}}$, i.e., one IG-Encoder is responsible for encoding in both tasks but it uses different prediction layers (Eq.~\ref{eq:bg_predict}) for different tasks.

     \item \textbf{IG-Encoder$_{\mathbf{M3S}}$} adopts the self-training approach proposed by Sun et al.~\cite{SunLZ20}. IG-Encoder$_\text{M3S}$ starts training over the labeled nodes, then assigns ``pseudo'' node labels to highly confident unlabeled nodes, and includes these nodes in the labeled set for the next round of training.

     \item \textbf{DCI}\footnote{\url{https://github.com/wyl7/DCI-pytorch}} decouples the representation learning phase and the detection phase as \ours. In the representation learning phase, it adopts the node clustering task as the self-supervised learning task to capture the intrinsic properties of the graph. In the detection phase, it adopts a simple linear mapping followed by a sigmoid activation function to predict the abnormal score of a node.

\end{enumerate}

IG-Encoder$_{\mathbf{MTL}}$ and IG-Encoder$_{\mathbf{M3S}}$ are IG-Encoders variations equipped with recently proposed self-supervised reinforcements for graph representation learning.

\end{document}